\documentclass[a4paper,fleqn,usenatbib]{mnras}

\usepackage[T1]{fontenc}
\usepackage{ae,aecompl}

\usepackage{graphicx}	
\usepackage{amsmath}	
\usepackage{amssymb}	

\title[PWN morphology simulations]{The imprint of pulsar parameters on the morphology of Pulsar Wind Nebulae}

\author[R. B\"uhler et al.]{
Rolf B\"uhler,$^{1}$\thanks{E-mail: rolf.buehler@desy.de}
Matteo Giomi$^{1}$
\\
$^{1}$DESY, Platanenallee 6, 15738 Zeuthen, Germany\\
}

\pubyear{2016}

\begin{document}
\maketitle

\begin{abstract}
The morphology of young Pulsar Wind Nebulae (PWN) is largely determined by the properties of the wind injected by the pulsar. We have used a recent parametrization of the wind obtained from Force Free Electrodynamics simulations of pulsar magnetospheres to simulate nebulae for different sets of pulsar parameters. We performed axisymmetric Relativistic Magnetohydrodynamics simulations to test the morphology dependence of the nebula on the obliquity of the pulsar and on the magnetization of the pulsar wind. We compare these simulations to the morphology of the Vela and Crab PWN. We find that the morphology of Vela can be reproduced qualitatively if the pulsar obliquity angle is $\alpha \approx 45^\circ$ and the magnetization of the wind is high ($\sigma_0 \approx 3.0$). A morphology similar to the one of the Crab Nebula is only obtained for low magnetization simulations with $\alpha \gtrsim 45^\circ$. Interestingly, we find that Kelvin-Helmholtz instabilities produce small scale turbulences downstream of the reverse shock of the pulsar wind.
\end{abstract}




\section{Introduction}
Most of the rotational energy lost by a pulsar is transferred to a relativistic particle wind. These particles are by numbers predominantly electrons and positrons (referred to together as electrons in the following).  The wind is thought to be cold, meaning that its thermal energy is much less than its bulk kinetic and magnetic energies. When the wind interacts with ambient material, the particles become isotropised and radiate \citep{Arons2012}. This is what is then seen as a Pulsar Wind Nebula (PWN) in the sky. To date, around 100 of these systems have been found \citep{Kargaltsev2015}. In the X-rays, several of them show a torus morphology, with a jet emerging perpendicular to it \citep{Kargaltsev2008}. 

Young PWN, with an age smaller than $\approx$10000 yrs, have not yet been distorted by the reverse shock of the stellar explosion \citep{Gaensler2006,Kargaltsev2015}. Their morphology is still closely related to the properties of the pulsar wind. Particularly in two cases, the Crab and Vela PWN, the plasma outflow can be resolved observationally in great detail, down to spatial scales of $\Delta r \lesssim 0.03$~ly. These systems therefore provide a test bed to study the behaviour of relativistic plasma, which is also of relevance for other non-thermal sources as Active Galactic Nuclei \citep{Netzer2014,Massaro2015} or Gamma-ray Bursts \citep{Gehrels2012,Berger2014}.

The properties of the pulsar wind -- as its magnetic field, particle and velocity distributions -- are not known with certainty today. However, over the past years there has been great progress in this respect. Several groups have performed Force Free Electrodynamics (FFE) \citep{Spitkovsky2006,Kalapotharakos2012,Tchekhovskoy2013} and Particle in Cell (PIC) simulations of pulsar magnetospheres  \citep{Philippov2014,Cerutti2015}. These simulations allow one to trace the properties of the wind out to several light cylinder radii $r_{lc}$. Due to the relativistic speed of the wind, it is expected that the wind does not have time to re-arrange itself on larger scales afterwards. In particular, the latitude dependence of its energy flux is expected to remain unchanged as the wind moves out into the nebula \citep{Tchekhovskoy2015}.

Recently, the first analytic parametrization of the latitude dependent luminosity of the pulsar wind has been derived from FFE simulations \citep{Tchekhovskoy2015}. The main parameter determining the wind properties is the obliquity angle between the pulsar spin axis and its magnetic moment $\alpha$. The latter can not be measured directly. Constraints from pulsar light curve modelling usually differ vastly between pulsar emission models \citep{Pierbattista2015}. The other important unknown parameter is the magnetization $\sigma$ of the wind. It is thought that most of the wind energy is in its magnetic fields ($\sigma \gg 1$) at $r_{LC}$ . When and where this energy is transferred to kinetic particle energy is not known. FFE simulations of pulsar magnetospheres do not include non-thermal particle acceleration, this question cannot be addressed by them.

In this paper we will study the dependence of the nebula morphology on $\alpha$ and $\sigma$. Both of these parameters strongly affect the forces acting on the wind plasma. They are therefore expected
to shape the morphology of the resulting nebula. We performed Relativistic Magnetohydrodynamic (RMHD) simulations to scan the parameter space of different obliquity angles and a high and low magnetization of the wind. RMHD simulations of PWN performed in the past have primarily focused on the Crab nebula \citep{Hester2008,Buhler2014}. Qualitatively, the toroidal structure and the jet are well reproduced in 2D axisymmetric simulations \citep{Komissarov2004,DelZanna2004,DelZanna2006,Volpi2009,Bucciantini2011}. Dynamically, the motion of thin filaments, so called ``wisps'' is also reproduced \citep{Camus2009}. Recently, the first 3D simulations of the Crab nebula showed that axisymmetric simulations overpredict the strength of the jet \citep{Porth2013,Porth2013b}. In addition, compared to 2D simulations, significant turbulence emerges far downstream of the wind termination shock in 3D. This potentially enhances the magnetic dissipation, allowing for larger magnetizations of the wind to reproduce the Crab's morphology. 

Unfortunately, performing several 3D simulations to scan the phase space of pulsar wind parameters is still computationally too expensive. We therefore performed 2D axisymmetric simulations. In contrast to most previous studies, we simulate both hemispheres. As was shown by \citet{Porth2013b}, this enhances the magnetic dissipation also in the axisymmetric case in the equatorial regions. Nevertheless, we will keep the axisymmetric limitation of our simulations in mind and will come back to it in the discussion of the simulation results in section \ref{sec:res}. 

We chose the length scales and spin-down power of the pulsar to values appropriate for the Vela PWN \citep{Pavlov2003,Durant2013}. To our knowledge this system has not been simulated in RMHD to date. We expect that apart from scaling factors, the PWN morphology does not depend strongly on this choice. Qualitatively, we expect the simulated morphologies to be similar also in other young PWN. We will use cgs units throughout, except for length scales, which for convenience will be given in light years.

\section{Pulsar Wind Nebula simulations}
In our simulations, a pulsar wind is injected into a homogeneous ambient medium at rest. The later has only the role to confine the pulsar wind. The regions of interest for the high-energy emission are in the inner nebula, close to the reverse shock of the pulsar wind. Once the pulsar wind has blown a sufficiently large bubble inside of its surrounding medium, the properties of the inner region do not depend strongly on the ambient environment. Instead, the inner nebula morphology is predominantly determined by the wind properties. As mentioned before, the latter are determined by the obliquity angle of the pulsars and by the wind magnetization. We performed six simulations with different combinations of these parameters shown in table \ref{tab:simpar}. The simulation setup will be described in more detail in the following.

\subsection{Simulation setup}
The simulations were run with the RMHD module of the \texttt{PLUTO}\footnote{\url{http://plutocode.ph.unito.it}} code version 4.2 \citep{Mignone2007}. The RMHD equations were evolved in time using an HLLC solver. We assumed a polytropic equation of state with an adiabatic index of $4/3$. We applied spherical coordinates, with a linear binning in the polar angle $\theta$ and a logarithmic binning in radius $r$. Adaptive Mesh Refinement (AMR) was used with four refinement levels, with a factor two in cell sizes between the levels \citep{Mignone2012}. The first grid was divided in 88 radial bins, from $r_{min} = 0.0002$~ly to $r_{max} = 1.4$~ly, and 32 polar angle bins. At a typical distance of the reverse shock of $r_{shock} = 0.05-0.2$~ly the resolution is $\Delta r_{shock} \approx (3-13) \times 10^{-4}$~ly at the highest AMR level. To speed up the simulations, AMR is not activated in the unshocked wind region for $ r > 0.001$~ly. Reflective boundary conditions were applied along the symmetry axis and continuous outflow boundary conditions were applied at $r_{max}$.  More details on the simulation parameters can be found in the \texttt{PLUTO} configuration files shown in Appendix~\ref{app:pluto}.

\begin{table}
	\centering
	\caption{Parameters for the performed simulations: the pulsar obliquity angle $\alpha$ and the wind magnetization before ($\sigma_0$) and after ($\bar{\sigma}$) the annihilation of the striped wind.  The last column indicates which source the simulated morphology shown in figures \ref{fig:emlsig} and \ref{fig:emhsig} resembles qualitatively (see text).}
	\label{tab:simpar}
	\begin{tabular}{ccccc} 
		\hline
		Sim. Nr. & $\alpha$& $\sigma_0$ & $\bar{\sigma}$ & Source\\
		\hline
		1 & 10$^\circ$ & 0.03 & 0.024 & \\
		2 & 10$^\circ$ & 3 & 1.5 &\\
		3 & 45$^\circ$ & 0.03 & 0.0043 &Crab\\
		4 & 45$^\circ$ & 3 & 0.12 & Vela\\
		5 & 80$^\circ$ & 0.03 & 0.00014 &Crab\\
		6 & 80$^\circ$ & 3 & 0.0038 &Crab\\
		\hline
	\end{tabular}
\end{table}

The pulsar wind was injected into an ambient medium at $r_{min}$. In order to speed up the simulations, we used a relatively thin ambient medium with a density of $\rho_{am}= 10^{-28}$~g~cm$^{-2}$. The work required to blow the pulsar wind bubble into the surrounding medium is thereby reduced. In order to further accelerate the simulations, the region within $r<0.06$~ly was filled with the unperturbed pulsar wind parameters at the start of the simulations.

We setup the pulsar wind parameters following the prescriptions given in \citet{Porth2013b}. The total energy flux density of the wind, $f_{tot}$, as a function of the polar angle $\theta$ and the obliquity angle of the pulsar is given by:
\begin{equation}
f_{tot}(r, \theta, \alpha) = \frac{L}{L' r^2} \times (g(\theta, \alpha) \times \sin^2 \theta  + d).
\end{equation}
The angular dependence of the wind is given by the function $g(\theta,  \alpha)$, and $d = 0.02$ is added for numerical reasons to avoid vanishing energy flux at the poles. $L'$ is a scaling factor that normalizes the total energy flux to the spin-down luminosity of the Vela pulsar $L = 6.9 \times 10^{36}$ ergs s$^{-1}$ \citep{Manchester2005}. The angular dependence of the wind is obtained by averaging the parametric solutions given by \citet{Tchekhovskoy2015} over the azimuth angle $\phi$:

\begin{equation}
\begin{aligned}
g(\theta, \alpha) =&  \langle (w_1(\alpha) b_1(\theta, \phi) + w_2(\alpha) b_2(\theta, \phi))^2 \rangle _\phi, ~\mathrm{with}\\
w1(\alpha) =& |1-2\alpha / \pi|,\\
w_2(\alpha) =& 1 + 0.17|\sin2\alpha| -  w1(\alpha),\\
  b_1(\theta, \phi) =& [1 + 0.02 \sin\theta_m + 0.22 (|\cos\theta_m|-1)\\
& - 0.07 (|cos\theta_m| -1)^4] \times sign \cos \theta,\\
 \theta_m(\theta, \phi) =& \arccos(\sin\alpha \sin\theta \cos\phi + \cos\theta \cos\alpha),\\
 b_2(\theta, \phi) =& \cos(\phi - 30^\circ)  \sin \theta
\end{aligned}
\end{equation}

\begin{figure*}
	\includegraphics[width=\textwidth]{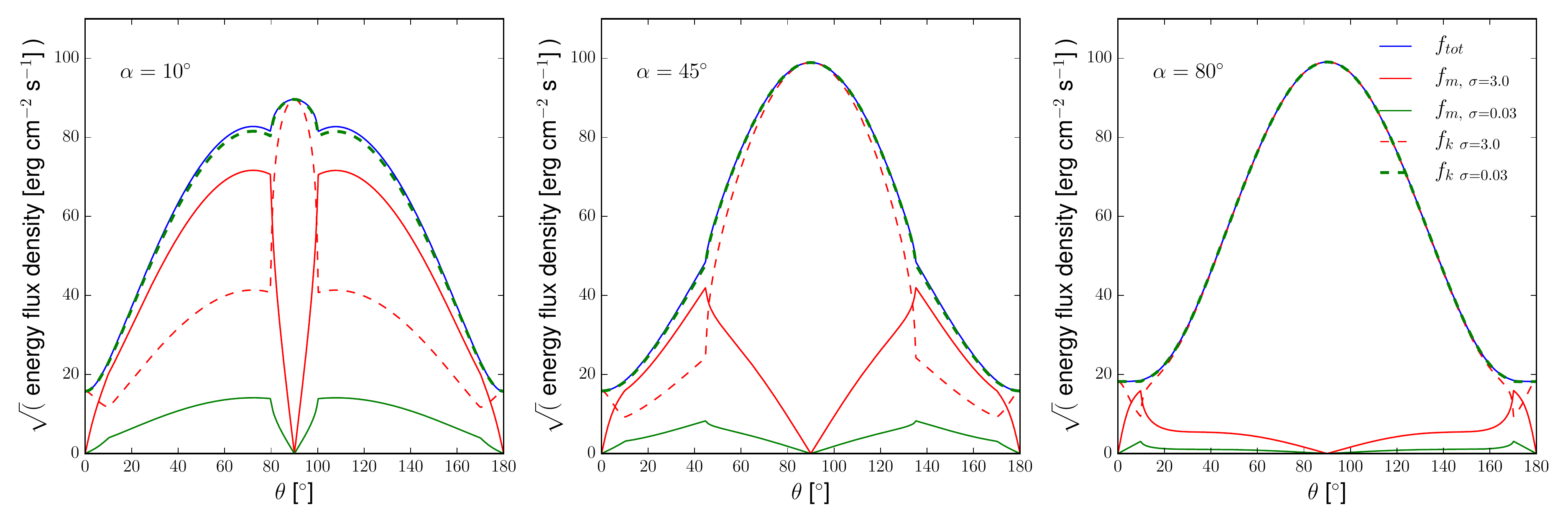}
    \caption{Components of the energy flux density of the simulated pulsar wind injected by the pulsar as a function of the polar angle $\theta$. The distributions are shown for different obliquity angles $\alpha$ and are normalized to a distance of 1~ly.}
    \label{fig:eflux}
\end{figure*}

The energy flux density is divided between a magnetic term $f_m$ and a kinetic term $f_k$:
\begin{equation}
\begin{aligned}
f_m (r,\theta, \alpha) =& \sigma(\theta) \frac{f_{tot}(r,\theta, \alpha)}{1 + \sigma(\theta)},\\
f_k (r,\theta, \alpha) =& \frac{f_{tot}(r,\theta, \alpha)}{1 + \sigma(\theta)},
\end{aligned}
\end{equation}
where $\sigma(\theta) \equiv f_m / f_k $ is the wind magnetization. To assure that the Poynting flux vanishes at the poles, the magnetization saturates at:
\begin{equation}
\tilde{\sigma} (\theta) = \left\{ \begin{array}{rl}
 (\theta/10^\circ)^2 \times \sigma_0 &\mbox{ $\theta \leq 10^\circ$} \\
  \sigma_0 &\mbox{ $\theta > 10^\circ$}
       \end{array} \right.
\end{equation}
, where $\sigma_0$ is one of the simulation parameters reported in table \ref{tab:simpar}. We assume that dissipation of magnetic fields in the striped wind zone changes the magnetization to
\begin{equation}
\sigma(\theta) = \frac{\tilde{\sigma_0} ~ \chi(\theta,\alpha)}{1 + \tilde{\sigma_0}(\theta)(1 - \chi(\theta,\alpha))},~\mathrm{where}
\end{equation}
\begin{equation}
\chi(\theta, \alpha) = \left\{ \begin{array}{rl}
 (2 \phi(\theta, \alpha) / \pi -1)^2 &\mbox{ $\pi/2-\alpha < \phi < \pi/2 + \alpha$ } \\
  1 &\mbox{ otherwise }
       \end{array} \right.
\end{equation}
with $\phi(\theta, \alpha) = \arccos(-\cot(\theta) \cot(\alpha))$. This assumes a perfect dissipation of the magnetic energy in opposite field lines in the stripped wind region. It is dynamically not important if the magnetic dissipation happens before the wind termination or right after it. Efficient dissipation after the wind termination has been shown in PIC simulation \citep{Sironi2011a}.  The full annihilation of opposite magnetic field lines in the stripped region implies that the average wind magnetization depends strongly on $\alpha$. In particular, for large values of $\alpha$ the average magnetization $\bar{\sigma}$ will be low, independently of $\sigma_0$, as can be seen in table \ref{tab:simpar}. The resulting energy flux distributions for the different simulations are shown in figure \ref{fig:eflux}.

The magnetic field in the simulations is assumed to be purely toroidal, and inverts its direction in the equatorial plane:
\begin{equation}
B_{\phi} = \pm \sqrt{4 \pi f_m(r, \theta)/c}
\end{equation}
Asymptotically one expects the poloidal and radial components of the magnetic field to decrease rapidly with distance with $r^{-2}$, while the toroidal component decreases with $r^{-1}$. For distances $ r \gg r_{LC}$ one therefore expects the toroidal component to dominate.  Observationally, the toroidal magnetic field structure is supported by the fact that polarization measurements of PWN show an aligned magnetic structure, which is perpendicular to the symmetry axis of the nebula \citep{Moran2014}.

One complication in RMHD simulations of PWN is that the wind is expected to have a Lorentz factor $\Gamma > 10^4$. However, the code becomes numerically unstable for $\Gamma > \Gamma_{max} \approx 10$. Fortunately, in PWN the global plasma dynamics are primarily determined by the total kinetic energy $f_k$ \citep{Porth2013b}. We therefore followed the approach taken by all previous RMHD simulation of PWN and compensated for the lower $\Gamma$ by increasing the plasma density:
\begin{equation}
\rho(r, \theta) = f_k (r, \theta)/(\Gamma_{max}^2 c^3)
\end{equation}

\begin{figure*}
	\includegraphics[width=\textwidth]{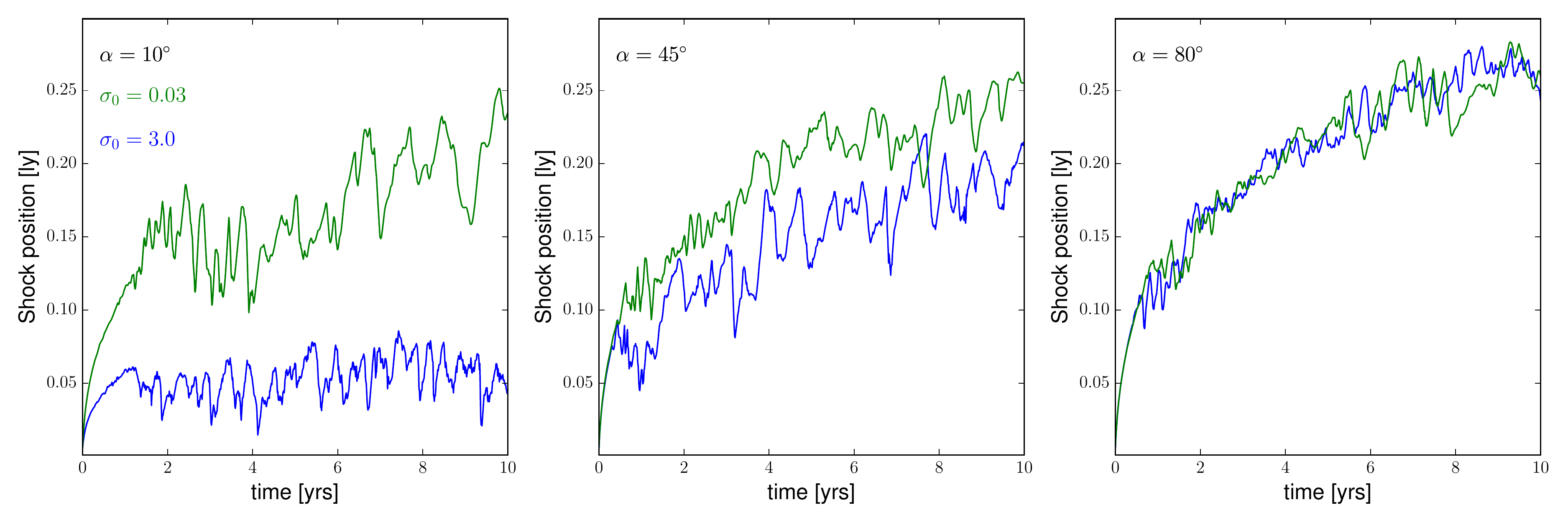}
    \caption{Position of the reverse shock of the pulsar wind as a function of time for different obliquity angles $\alpha$ and magnetizations $\sigma_0$. The shock position has been averaged over 10$^\circ$ around the equator.}
    \label{fig:shock}
\end{figure*}

\begin{figure*}
	\includegraphics[width=\textwidth]{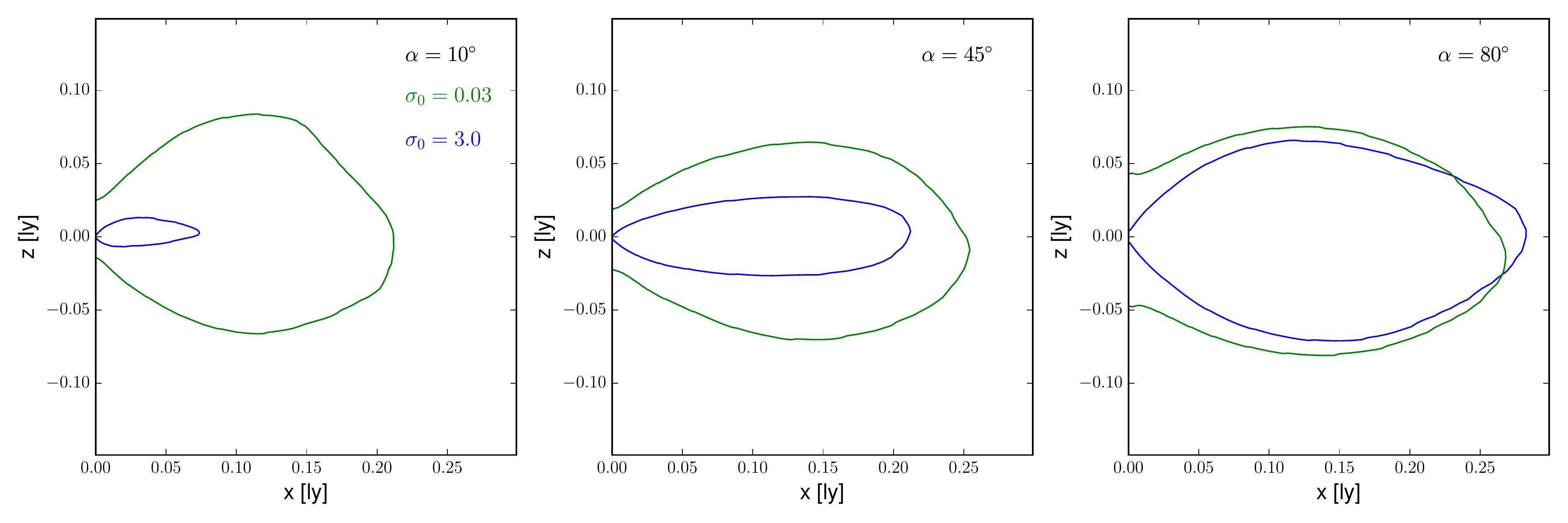}
    \caption{Shape of the reverse shock of the pulsar wind as a function of time for different obliquity angles $\alpha$ and magnetizations $\sigma_0$. The shock position has been averaged over time from $t_{sim} = 8$~yrs to $t_{sim} = 10$~yrs.}
    \label{fig:shockshape}
\end{figure*}

The simulations were performed over a time period of $t_{sim} = 10$~yrs. After this time, the PWN has reached the self similar expansion phase, where spatial scales increase approximately linearly \citep{Komissarov2004}. This can be seen in fig. \ref{fig:shock} for the position of the reverse shock: some initial oscillations are followed by a rapid expansion. Around $t_{lin} \gtrsim 6$~yrs, the reverse shock position begins to expand more slowly and approximately linear in time for all simulations. The average shape of the shock surface as a function of latitude is shown in figure \ref{fig:shockshape}.

\subsection{Radiative model}

Synchrotron emission of relativistic electrons energetically dominates the radiative output of PWN. Over a wide range in frequency $\nu$, non-thermal photon spectra are observed. The photon flux is typically well described by a power-law function $f(\nu) \propto \nu^{-\lambda}$. A typical value for the photon index in the X-rays is $\lambda \approx 2$, which we will assume throughout.

In order to properly describe the synchrotron emission of PWN, the energy distribution of electrons as a function of position and energy needs to be known. In RMHD emission models of the Crab nebula, particles are typically injected at the wind termination shock. However, it is not known what the latitude dependence of the injected spectrum is \citep{Olmi2014}. Furthermore, it is well possible that non-thermal particle acceleration also happens in the body of the nebula, far downstream of the wind termination shock \citep{Gallant1994,Amato2006,Bucciantini2011}. Given these uncertainties, we proceed using a simpler approach. Following \citet{Komissarov2004} and \citet{DelZanna2004}, we assume that the photon emissivity $\epsilon$ depends primarily on the Doppler factor $D = 1 / \Gamma ( 1 - \boldsymbol \beta \textbf{n} )$ and magnetic field $B_p'$ perpendicular to the line of sight $\textbf{n}$ in the plasma rest frame:

\begin{equation}
\epsilon \propto D^{\lambda + 2} B_p'^{\lambda +1}, \mathrm{where}
\end{equation}
\begin{equation}
B_p' = \sqrt{ B^2 - D^2 (\textbf{B n})^2 + 2 \Gamma D(\textbf{B n})(\boldsymbol \beta \textbf{B})} / \Gamma
\end{equation}

and $\boldsymbol \beta$ is the velocity vector normalized to the speed of light. Previous studies have shown that $D$ and $B_p'$ are the most important parameters in determining the synchrotron emissivity \citep{DelZanna2006,Camus2009,Bucciantini2011,Porth2013b}. We will therefore use the prescription above to gain a qualitative picture of the PWN morphology and to identify the most likely radiation regions. After having described the simulation procedure, we proceed to discuss their results.

\section{Results}
\label{sec:res}

In the following, we will focus our discussion on the snapshot at the end of the simulations at $t_{sim} = 10$~yrs. The nebula is already close to a self similar expansion at this time.

\subsection{The plasma flow}
\label{sec:flow}

\begin{figure*}
\begin{center}
	\includegraphics[width=0.9\textwidth]{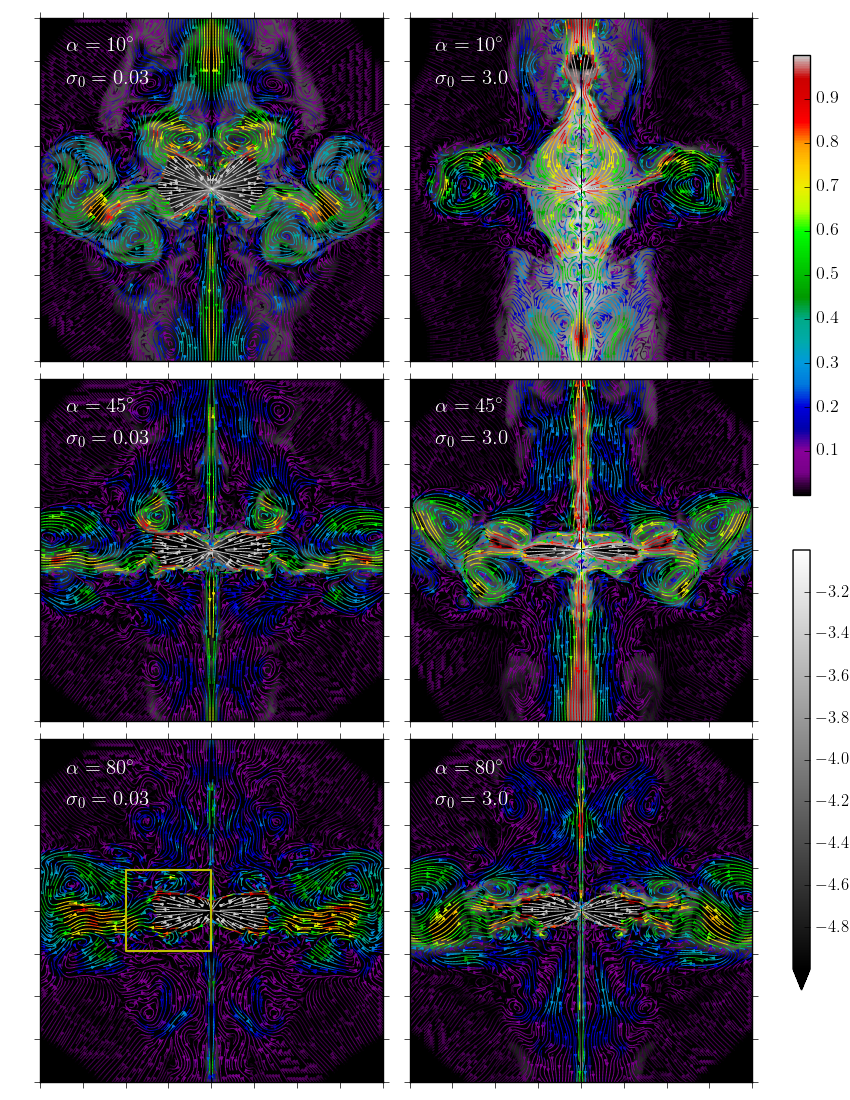}
 \end{center}
     \caption{Streams of the plasma flow for different simulations at a time $t_{sim} = 10$~yrs. The colours show the plasma speed in units of c. The background image shows the logarithm of the magnetic field strength in a grey scale. The spatial extend of each panel is 1.6~ly on each side. A zoom of the region indicated by the yellow box in the lower left panel is shown in figure \ref{fig:streamszoom}. }
    \label{fig:streams}
\end{figure*}

The global plasma flow patterns of the different simulations are shown in figure \ref{fig:streams}. Common patterns are observed for all simulations: the pulsar wind is decelerated abruptly at an oblate reverse shock. Plasma at higher latitudes is transported to the polar regions due to the magnetic hoop stress and a jet emerges. In the equatorial region, plasma continues to move radially outward, creating a torus in the equatorial plane. The torus region is highly turbulent and the radial flow pattern is lost. The polar flow becomes stronger compared to the equatorial one with increasing average wind magnetization. As expected, the most relevant parameter is $\bar{\sigma}$ and not $\sigma_0$. This can be seen clearest for the $\alpha=80^{\circ}$ case, where the flow patterns are very similar for $\sigma_0= 0.03$ and $\sigma_0= 3.0$.

\begin{figure*}
\begin{center}
	\includegraphics[width=1.0\textwidth]{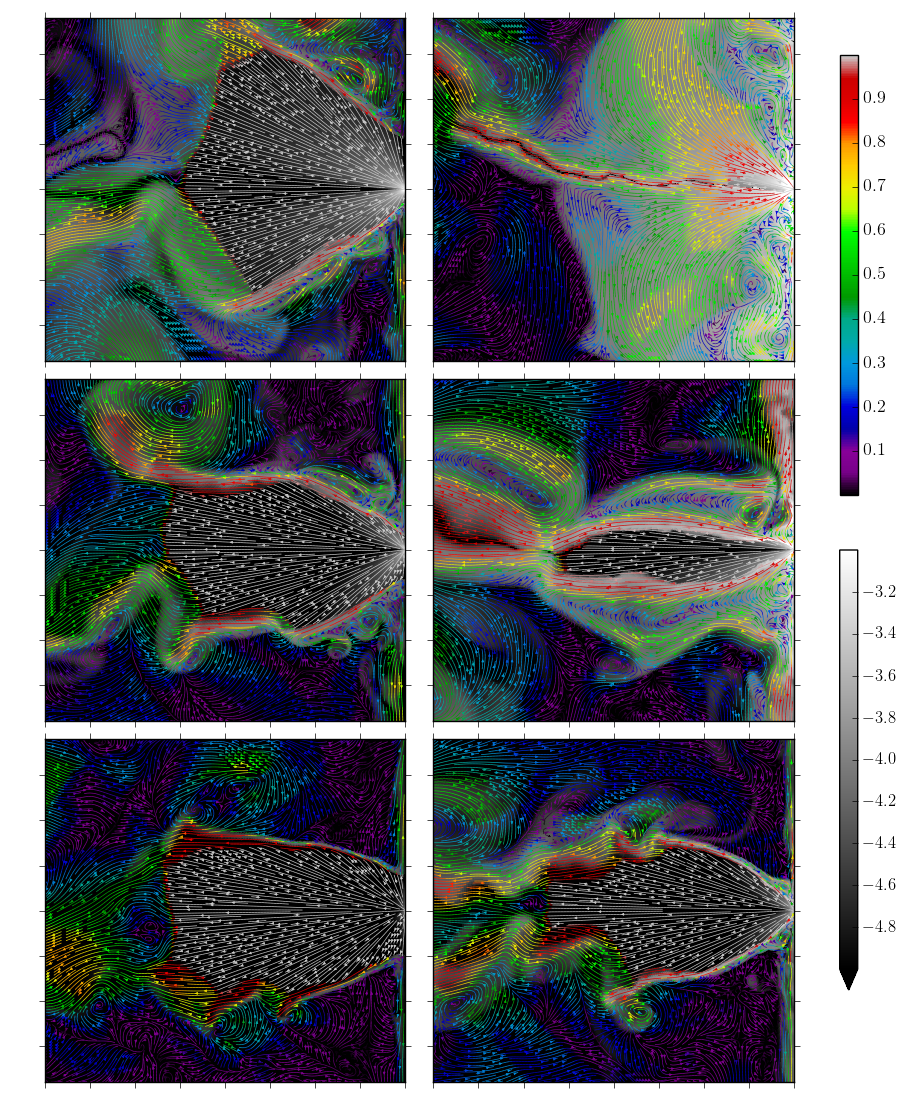}
 \end{center}
     \caption{As figure \ref{fig:streams}, but zoomed in on the inner nebula region. The shown region is indicated in the lower left panel of figure \ref{fig:streams}. It extends from $x=-0.4$~ly to $x=0$~ly  and $z= -0.19$~ly to $z=0.19$~ly.}
    \label{fig:streamszoom}
\end{figure*}

A zoom in on the flow pattern in the region close to the reverse shock is shown in figure \ref{fig:streamszoom}. The region before the reverse shock, referred to as ``wind region'' in the following, becomes smaller with increasing $\bar{\sigma}$. This can also be seen in the shock position at different latitude angles shown in figure \ref{fig:shockshape}. The wind region also becomes more oblate with increasing $\bar{\sigma}$. This is expected, as the increased hoop stress results in increasing pressure in the polar regions  \citep{Lyutikov2015}.  Despite these dependencies, the flow patterns are similar. The exception is the simulation of high magnetization and a small obliquity angle ($\sigma_0 = 3.0$ and $\alpha = 10^{\circ}$). The size of the wind region is greatly reduced and the polar flow greatly enhanced compared to the other simulations. 

It is apparent from figure  \ref{fig:streamszoom} that the degree of plasma turbulence increases with decreasing $\bar{\sigma}$. The flow is very regular for the $ \sigma_0 = 3.0$ and $\alpha=10^{\circ}$ simulation. There is also a regular plasma flow in the downstream of the reverse shock for $ \sigma_0$ and $\alpha=10^{\circ}$. These are the two simulations with the largest $\bar{\sigma}$. In contrast, all other simulations show a high degree of turbulence, which emerges almost directly downstream of the termination shock.

\begin{figure}
	\includegraphics[width=\columnwidth]{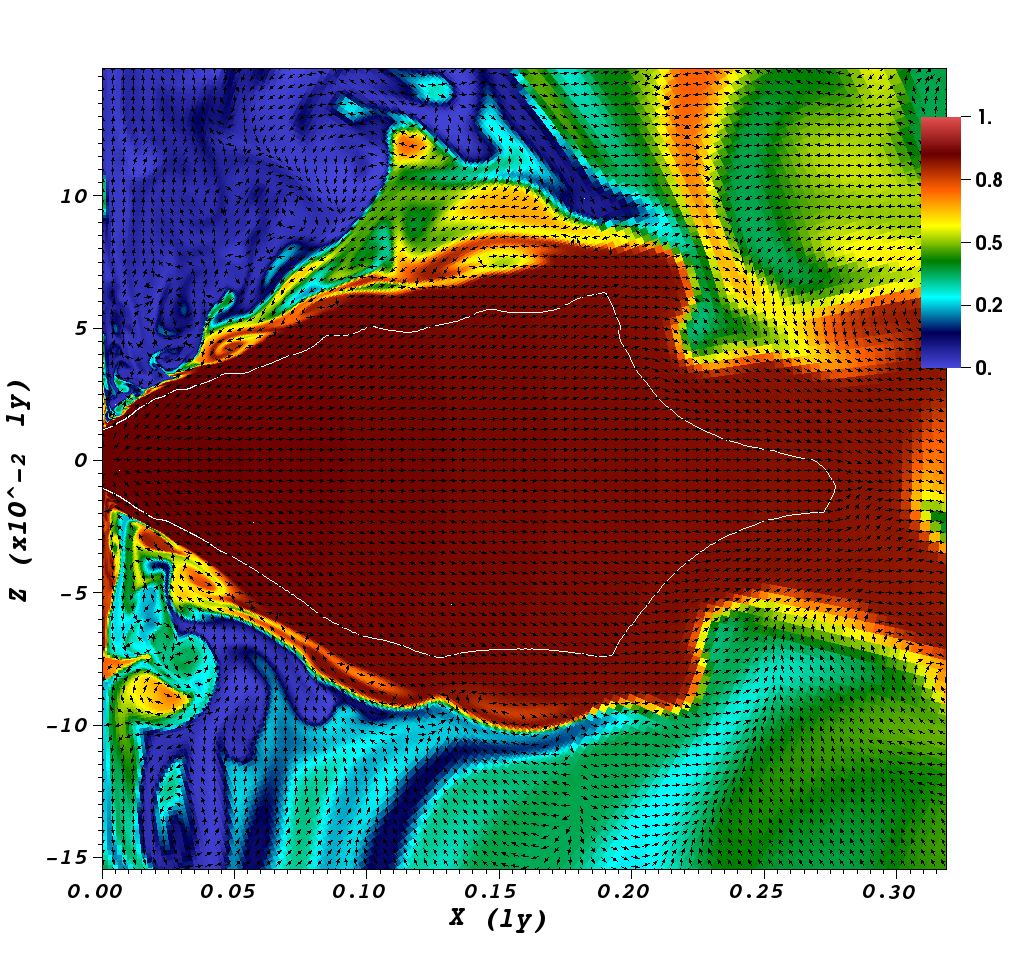}
	\includegraphics[width=\columnwidth]{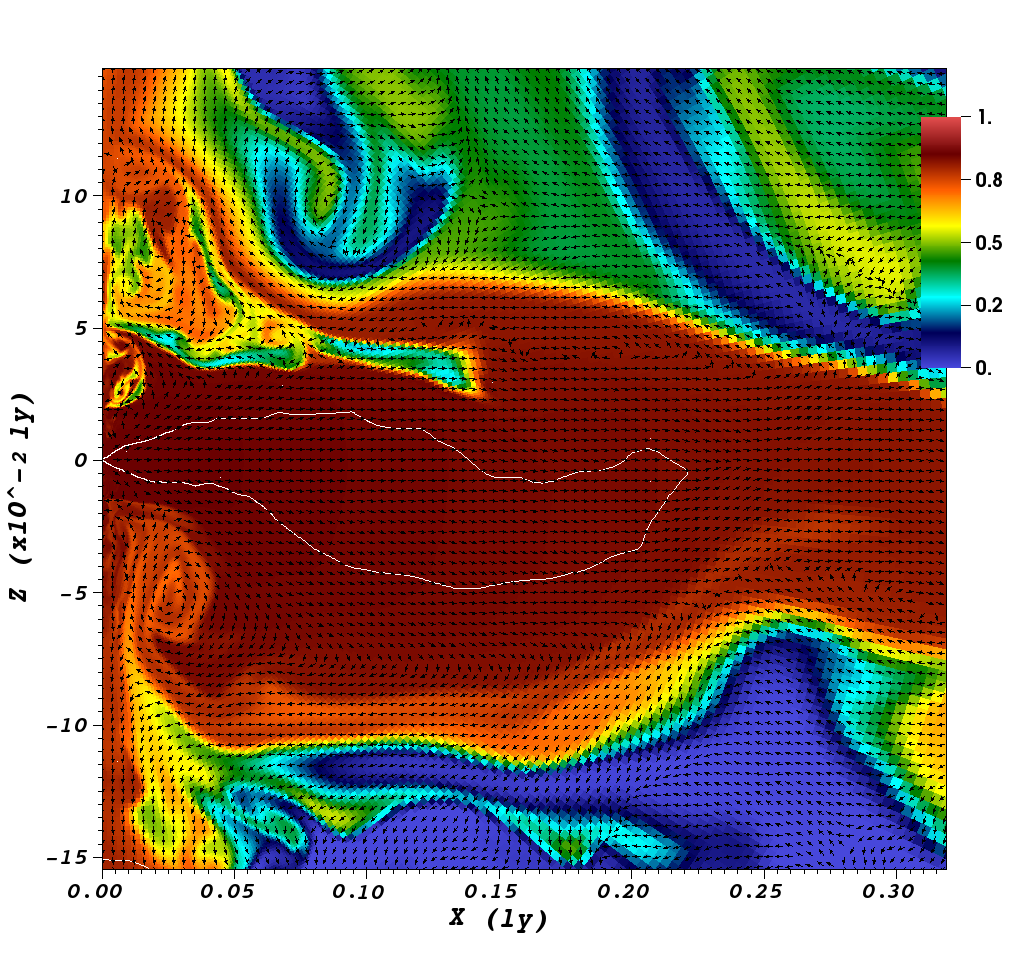}
    \caption{Region around the reverse shock for the simulations of $\alpha=45^{\circ}$ with $ \sigma_0 = 0.03$ (top panel) and  $\sigma_0 = 3$ (bottom panel) at a time of $t_{sim} = 8.5$~years. The colors show a tracer which indicates the time when the plasma was injected into the nebula in units of 10 years. Movies showing the emergence of the instability also for the other simulations are available online \citep{Buehler2016}. Arrows show the direction of the plasma flow. The white line marks the reverse shock position at the point $\Gamma  = 8 $. KH instabilities can be seen along the high and low latitude oblique shock (see text). }
    \label{fig:kh}
\end{figure}

Interestingly, loop like patterns from magnetic Kelvin-Helmholtz (KH) instabilities are observed downstream of the shear flow regions. An example is shown in figure \ref{fig:kh}. As the shear flow is reduced for high magnetizations, KH loops are predominantly observed in the low-sigma simulations.
KH instabilities in PWN have been studied by \citet{Bucciantini2006} for local features within nebulae. To our knowledge, KH instabilities have not been reported in previous publications of global PWN simulations. However, its emergence had already been seen in the simulations discussed in the PhD thesis of \citet{Camus2009a}. The reason that most previous simulations did not reveal this instability is likely that their spatial resolution was a factor $>5$ larger compared to the ones presented here \citep{DelZanna2006,Camus2009,Bucciantini2011}.  As will be discussed in more detail in section \ref{sec:crab}, the presence of the KH instability could lead to interesting radiative signatures.

\subsection{Synchrotron emission maps}
\label{sec:maps}

\begin{figure*}
\begin{center}
	\includegraphics[width=1.0\textwidth]{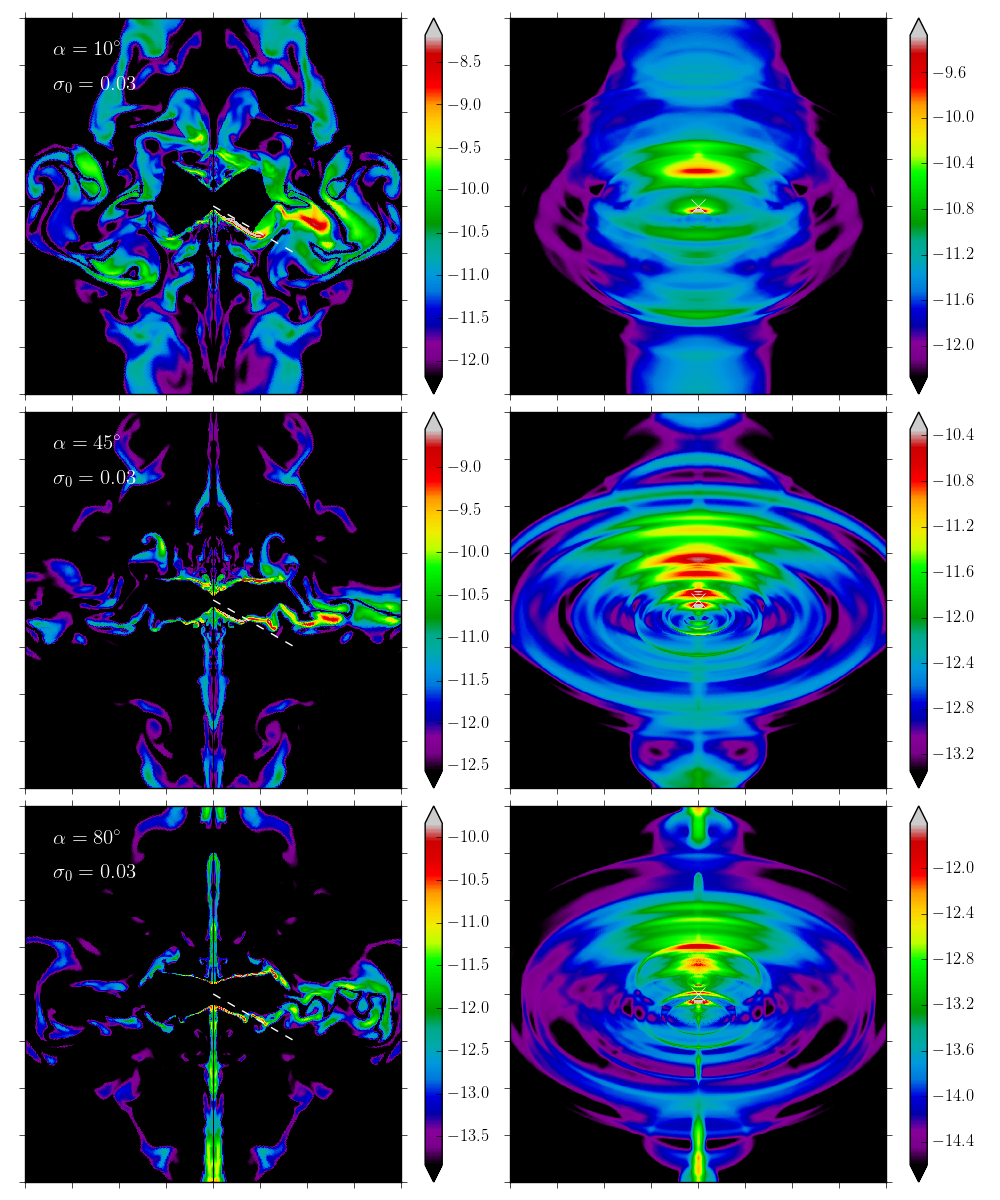}
 \end{center}
     \caption{Emission maps calculated for simulations of low magnetization $\sigma_0 = 0.03$ in arbitrary units. The left panels show a slice through the $y = 0$ plane. The right panels show the integrated emission in the line of sight for an observed at a viewing angle $\theta_{view} = 120^{\circ}$, indicated by the dashed white line in the left panels. The spatial extend of each panel is 1.6~ly on each side. }
    \label{fig:emlsig}
\end{figure*}
\begin{figure*}
\begin{center}
	\includegraphics[width=1.0\textwidth]{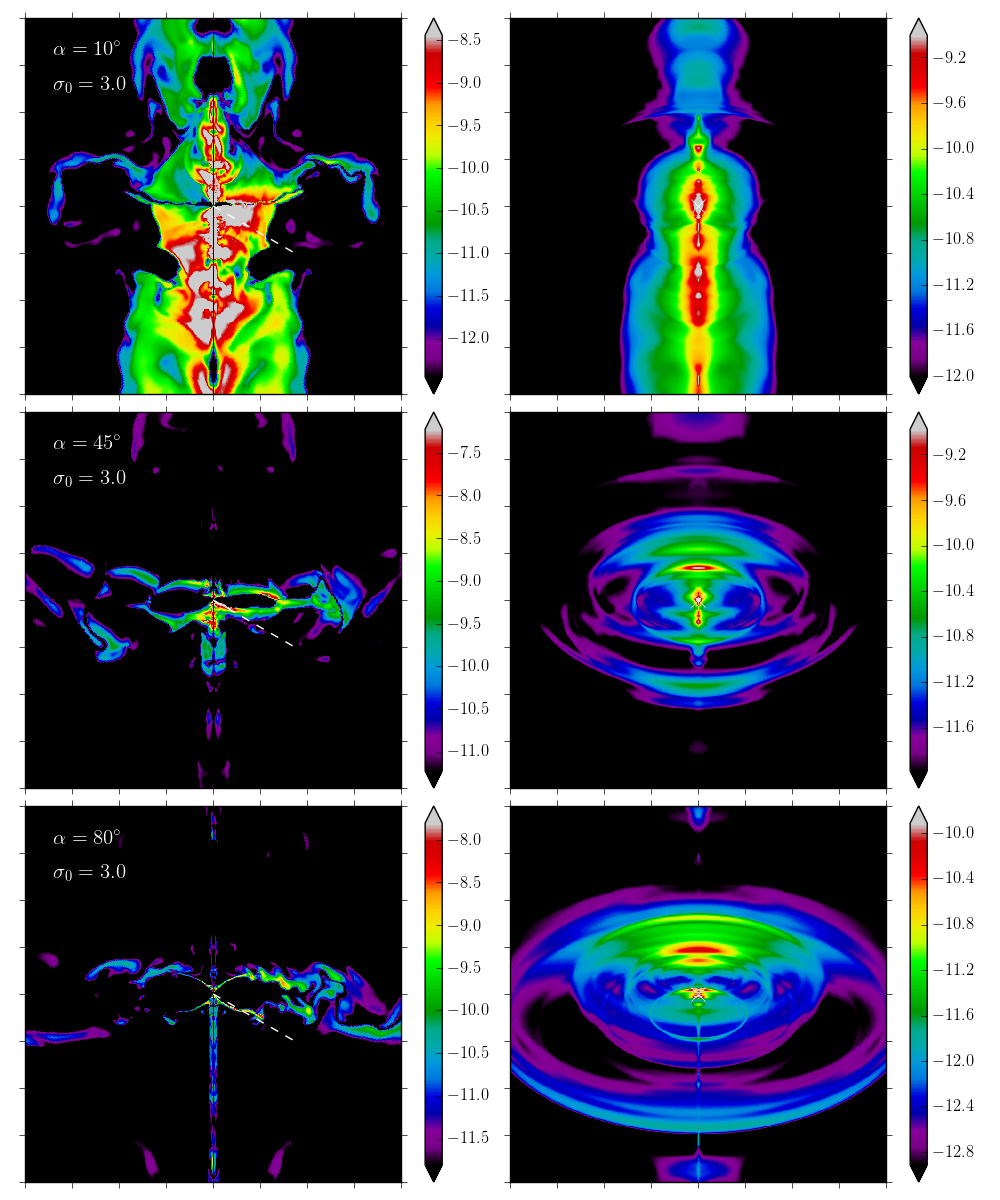}
 \end{center}
     \caption{As figure \ref{fig:emlsig}, but for simulations of high magnetization $\sigma_0 = 3.0$.}
    \label{fig:emhsig}
\end{figure*}

In this section, we will look at the morphology expected from the synchrotron emission of the plasma flow described in the previous section. The emission of the inner regions of PWN is very anisotropic due to the strong Doppler boosting. The viewing angle therefore plays an important role in determining the observed morphology. For our discussion, we will assume a viewing angle of $\theta_{view} = 120^{\circ}$. This is approximately the angle under which we observe the Vela and Crab nebulae from Earth \citep{Weisskopf2000,Helfand2001,Ng2004,Weisskopf2012}. These sources will be discussed in more detail in sections \ref{sec:vela} and \ref{sec:crab}. Maps for viewing angles of $\theta_{view} = 90^{\circ}$ and $\theta_{view} = 150^{\circ}$ are shown in appendix \ref{app:moremaps}.

The synchrotron maps for simulations with $\sigma_0=0.03$ are shown in figure \ref{fig:emlsig}. All of them show a torus, which comes from the equatorial region of the PWN. The synchrotron maps for $\alpha = 45^{\circ}$ and $\alpha = 80^{\circ}$ are qualitatively similar, with wider rings than the ones seen in simulation with $\alpha = 10^{\circ}$. The reason for this is that in the latter case more of the emission comes from higher latitudes. This is a result of the higher average magnetization and higher wind energy flux for $\alpha = 10^{\circ}$ (see figure \ref{fig:eflux}).

Figure \ref{fig:emhsig} shows the synchrotron maps calculated for the simulations done with $\sigma_0=3.0$. In this case, the difference between different obliquities is stronger. The main reason for this is that for $\alpha = 10^{\circ}$ and $\alpha = 45^{\circ}$ the average magnetization is close to unity. This results in elongated synchrotron nebulae. In the case of $\alpha = 10^{\circ}$, the torus is barely visible anymore as the emission is dominated by the higher latitudes. The jet is also clearly visible in this case.

Before we continue to confront these simulations with observations, we would like to recall two caveats: \textbf{(1)} Our simulations are axisymmetric. 3D simulations have shown that axisymmetry is a good approximation close to the wind reverse shock, in particular in the equatorial regions  \citep{Porth2013b}.  However, at high latitudes and further away from the reverse shock, 3D simulations show significantly different emission patterns. It follows from this that emission associated to the equatorial region in the inner nebula can be expected to be more robust. Polar regions on the other hand, and in particular the jet, are less trustworthy and will therefore not be a focus of discussion. \textbf{(2)} The emission model we apply here ignores spatial differences in the particle distribution function. The lack of strong spectral variation in the emission from the inner regions of PWN indicates that this is likely a good approximation \citep{Mori2004a}. Nevertheless, this is certainly oversimplified; e.g. individual structures in the Crab nebula are known to be visible only at particular wavebands \citep{Hester2002}. Keeping these caveats in mind, we will proceed to compare our simulation results to the Vela and Crab nebulae.

\subsection{Vela Pulsar Wind Nebula}
\label{sec:vela}

The Vela PWN is embedded in the Vela Supernova Remnant, also known as G263.9-3.3. Several regions of non-thermal emission are known within this remnant. The brightest one in radio is labelled Vela-X. The Vela pulsar is located in Vela-X \citep{Horns2006,DeJager2008,Abramowski2012,Grondin2013}. It is one of the closest and brightest gamma-ray pulsar known to date. Its distance has been determined via parallax to $936^{+62}_{-55}$~ly \citep{Caraveo2001,Dodson2003}. Due to its proximity the Vela pulsar might contribute significantly to the local cosmic-ray electron flux \citep{Hinton2011}. In the region surrounding the pulsar, the Chandra X-ray Observatory revealed a double ring structure, which is likely related to the reverse shock of the pulsar wind \citep{Helfand2001,Pavlov2003,Durant2013}. It is this innermost emission which we aim to reproduce in the simulations here. Unfortunately, the observational data is restricted to the X-ray band, as the ring structure has not been detectable so far at other wavebands \citep{Moran2014,Marubini2015}.

The three dimensional structure of the Vela rings has been interpreted as two equal torii, which are ontop of each other  (\citet{Helfand2001}, for an high contrast image of the Vela rings see figure 2 in \citet{Pavlov2003}). This results in a morphology which is more elongated along the symmetry axis compared to the Crab Nebula. The rings were also found to be closer to the pulsar as the innermost ring in the Crab nebula.  It was suggested that this might be due to a larger wind magnetization of order unity \citep{Helfand2001}. Indeed, our simulations confirm that solutions with higher magnetization resemble the ring structure better. Solutions with a lower $\bar{\sigma}$ -- as all simulations with $\sigma_0=0.03$ and both simulations with $\alpha=80^{\circ}$ -- are to wide and the torii are too narrow. Interestingly, several bright small-scale structures are found along the symmetry axis of the nebula. These features are related to the highly beamed right downstream of the reverse shock. Such features are also observed in the X-ray data \citep{Levenfish2013}. 

The best morphology agreement is found for the $\alpha=45^{\circ}$ and $\sigma_0=3.0$ simulation. In this simulation rings of similar size originate downstream of the reverse shock. One lower ring from the equatorial regions right behind the wind termination and an upper ring from the torus. In contrast, all other solutions result in rings of increasing size as one moves away from the pulsar. The exceptions are the simulations for $\alpha = 10^{\circ}$. However, for a $\sigma_0 = 3.0$ the nebula is far too elongated and for the case of $\sigma_0 = 0.03$ no double ring structure is obtained. The value of $\alpha=45^{\circ}$ agrees well with the value of $\alpha=53^{\circ}$ inferred from polarization measurements of the pulsar profile \citep{Johnston2005}. Models of the pulsed gamma-ray emission typically give higher values $\alpha=62^{\circ}$~--~$75^{\circ}$ \citep{Abdo2010a}.

\subsection{Crab Pulsar Wind Nebula}
\label{sec:crab}

The Crab is the most studied PWN. Its torus and jet structure can be observed in great detail from the radio to X-ray band \citep{Hester2008,Buhler2014}. It has therefore been the primary target for RMHD studies of PWN. Several authors have qualitatively reproduced the morphology of the inner nebula in axisymmetric simulations assuming a low wind magnetization $<\sigma> \approx 0.01$ \citep{Komissarov2004,DelZanna2006,Volpi2009,Camus2009,Porth2013b}. We confirm this findings in our simulations. A good agreement is found for case of $\alpha = 45^\circ$ and $\sigma_0=0.03$ and for both simulations with $\alpha = 80^\circ$. Also in agreement with previous studies, a small bright feature is found just below the pulsar position \citep{Lyutikov2015,Yuan2015}. This ``inner knot'' was proposed to be the site of the recently discovered gamma-ray flares \citep{Buehler2012,Mayer2013}. However, no observational evidence for this has been found to date \citep{Rudy2015}.  
 
It is puzzling, that the innermost ring of the Crab nebula observed in X-rays does not show a brightness profile as expected from Doppler beaming. The back side of the inner ring has a brightness which is comparable to its front side. In RMHD simulations the ring is found to be much fainter than in observations \citep{Porth2013b}. In addition, the ring is composed of a series of knots, in contrast to the smooth profiles found in the simulations. Our simulations show a similar disagreement. However, the observed KH instabilities at the shear flow downstream of the reverse shock lead to Doppler boosted emission regions also on the back side of emission rings close to the reverse shock. This results in a brighter emission from the receding part of the flow than expected in radial flow models. Our simulations show that this effect is not strong enough to result in a ring of equal brightness. However, the increased turbulence in the KH loops might trigger increased particle acceleration via magnetic reconnection \citep{Cerutti2013,Sironi2014} and magnetoluminescense \citep{Blandford2014,East2015,Nalewajko2016,Yuan2016,Lyutikov2016}. Stochastic Fermi acceleration might also occur \citep{Rieger2006}. In either case, fresh injection of high energy particles in back-flowing plasma could lead to a decreased Doppler asymmetry only for the highest synchrotron frequencies. Indeed, the inner ring is only observed in X-rays and has not been detected at lower frequencies to date \citep{Hester2002}. As the development of the KH instability is reduced for higher plasma magnetization, the latter could be constrained if this interpretation is correct.  Testing this idea quantitatively requires dedicated simulations which include particle acceleration, which is beyond the scope of this paper.

\section{Summary and outlook}
\label{sec:summary}

We have performed axisymmetric RMHD simulations of PWN to scan the parameter space for different pulsar wind properties. We have simulated the wind emerging for different pulsar obliquities, which has recently been derived from FFE simulations \citep{Tchekhovskoy2015}. In addition we have tested different wind magnetizations. In general, we find that the average wind magnetization is the most important parameter in determining the PWN morphology. The main effect of increasing obliquity angle is to increase the size of the striped wind region, where we have assumed a perfect dissipation of opposite magnetic field lines. We found that the wind region upstream of the reverse shock is smaller in size and becomes more oblate with increasing $\bar{\sigma}$.

With the exception of the wind morphology for $\alpha=10^{\circ}$ and $\sigma_0=3.0$, all simulations showed a torus in their emission maps, which emerges from the equatorial region downstream of the reverse shock. We have compared the morphologies of the different simulations to the Vela and Crab PWN. For Vela, we found that the simulation with the parameters $\alpha=45^{\circ}$ and $\sigma_0=3.0$ gives the best match to the observed morphology. For the Crab nebula, all simulations with a low $\bar{\sigma}$ match the observed morphology ($\alpha=10^{\circ}$ and $\sigma_0=0.03$ ; $\alpha=45^{\circ}$ and $\sigma_0=0.03$ ; $\alpha=80^{\circ}$ for $\sigma_0=0.03$ and $\sigma_0=3.0$ ).

We found that, particularly for low magnetizations, KH instabilities develops at the downstream at the shear flow of the reverse shock. The KH loops have the effect to increase the emission of the receding side of the nebula compared from what is expected from Doppler boosting of a radial flow. We suggest that this effect might help to explain that the innermost ring observed in X-rays in the Crab nebula has almost constant brightness.

We have pointed out the caveat that these conclusions rely on axisymmetric simulations. It would be desirable to confirm these findings with 3D simulations in the future. For a more quantitative comparison of observations and simulations, it will also be important to include a model for particle acceleration in the simulations. From the observational side, the detection of the rings observed in the Vela PWN outside of the X-ray band would be crucial to constrain the electron energy distribution. Taken together these steps provide the prospects in understanding the plasma flow quantitatively in PWN. This would be the first time this is achieved for relativistic plasmas flows and would likely have implications also for other sources as GRBs or AGN.

\section*{Acknowledgements}

We thank the referee Oliver Porth for very helpful comments and discussions. We believe they greatly improved the quality of the article. We also want to thank Andrea Mignone and Claudio Zanni very much for their help with the \texttt{PLUTO} code.


\bibliographystyle{mnras}
\bibliography{PWNsim} 

\clearpage
\appendix

\section{\texttt{PLUTO} configuration file}
\label{app:pluto}

The relevant parts of the \texttt{pluto.ini} file are listed below.
\\\\
\fbox{
  \parbox{\linewidth}{
\texttt{[Grid]\\
X1-grid    1    0.00002         88   l+   0.14\\
X2-grid    1    0.0             32   u    3.14159265358979\\
X3-grid    1    0.0             1    u    1.0\\
}

\texttt{[Chombo Refinement]\\
Levels           4\\
Ref\_ratio        2 2 2 2 2\\
Regrid\_interval  2 2 2 2\\
Refine\_thresh    0.8\\
Tag\_buffer\_size  3\\
Block\_factor     8\\
Max\_grid\_size    128\\
Fill\_ratio       0.4\\
}

\texttt{[Solver]\\
Solver         hllc\\
}

\texttt{[Boundary]\\
X1-beg        userdef\\
X1-end        outflow\\
X2-beg        axisymmetric\\
X2-end        axisymmetric\\
X3-beg        outflow\\
X3-end        outflow\\
}
  }
}
\\\\
The relevant parts of the \texttt{definitions.h} file are listed below.
\\\\
\fbox{
  \parbox{\linewidth}{
\texttt{\#define  PHYSICS                 RMHD\\
\#define  DIMENSIONS              2\\
\#define  COMPONENTS              3\\
\#define  GEOMETRY                SPHERICAL\\
\#define  RECONSTRUCTION          LINEAR\\
\#define  TIME\_STEPPING           RK2\\
}

\texttt{\#define  EOS                     IDEAL\\
\#define  ENTROPY\_SWITCH          CHOMBO\_REGRID\\
}

\texttt{\#define  SHOCK\_FLATTENING          MULTID\\
\#define  LIMITER                   DEFAULT\\
\#define  RECONSTRUCT\_4VEL          YES\\
\#define  CHOMBO\_LOGR               YES\\
\#define  RMHD\_FAST\_EIGENVALUES     YES\\
}
}}

\section{Additional maps}
\label{app:moremaps}
\clearpage

\begin{figure*}
\begin{center}
	\includegraphics[width=1.0\textwidth]{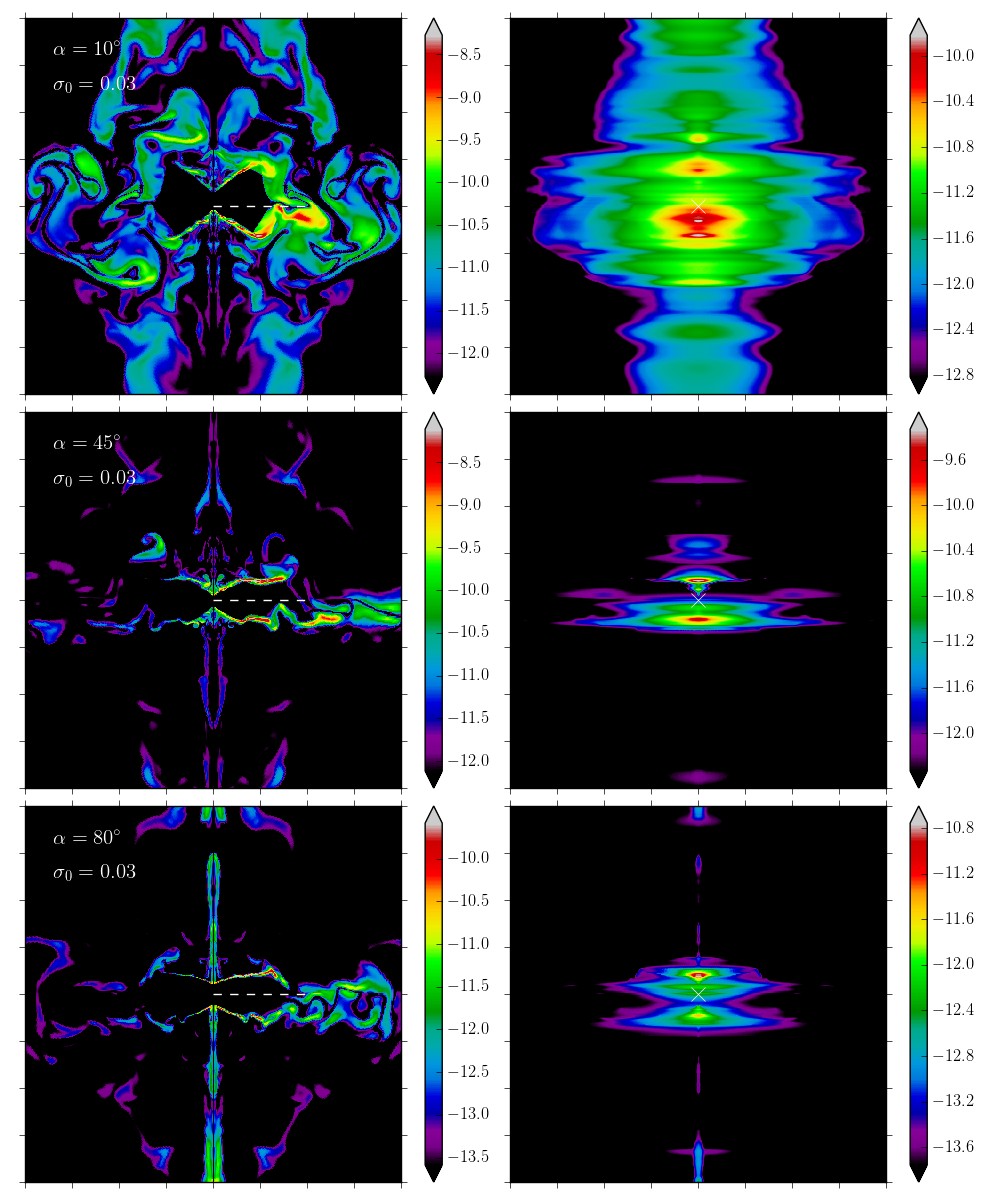}
 \end{center}
     \caption{As figure \ref{fig:emlsig}, but for a viewing angle $\theta_{view} = 90^{\circ}$.}
    \label{fig:emlsig90}
\end{figure*}
\clearpage
\begin{figure*}
\begin{center}
	\includegraphics[width=1.0\textwidth]{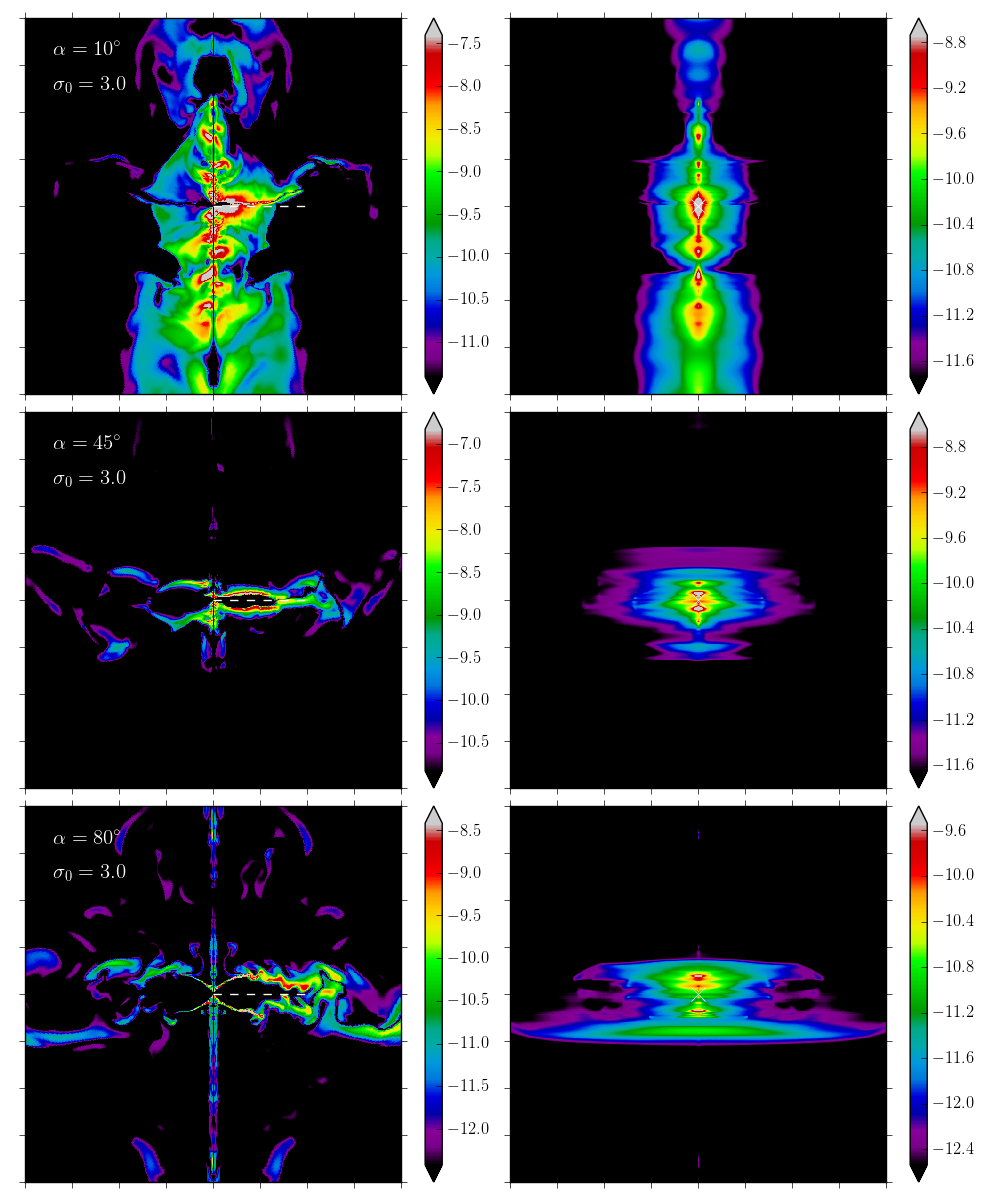}
 \end{center}
     \caption{As figure \ref{fig:emhsig}, but for a viewing angle $\theta_{view} = 90^{\circ}$.}
    \label{fig:emhsig90}
\end{figure*}
\clearpage
\begin{figure*}
\begin{center}
	\includegraphics[width=1.0\textwidth]{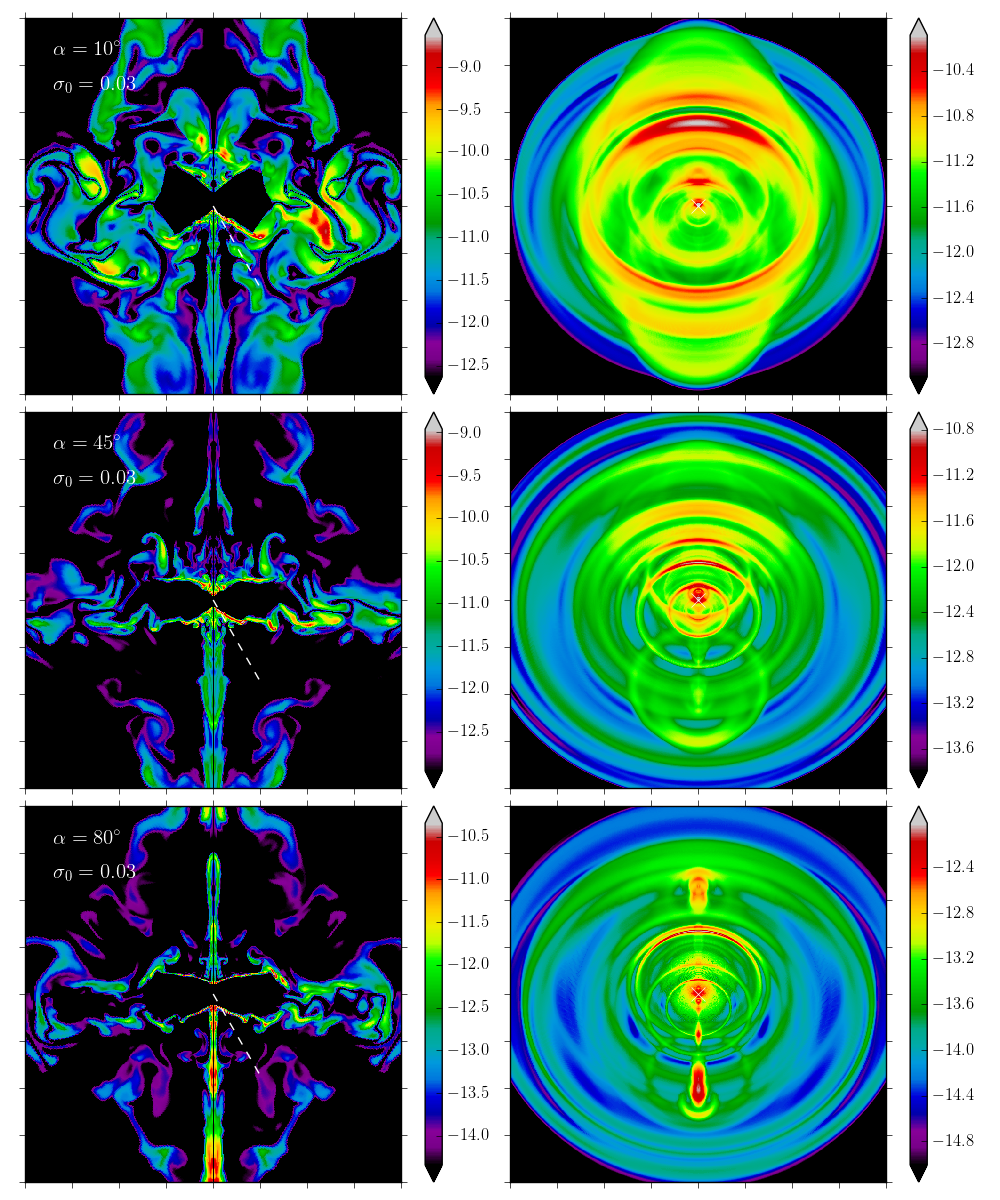}
 \end{center}
     \caption{As figure \ref{fig:emlsig}, but for a viewing angle $\theta_{view} = 150^{\circ}$.}
    \label{fig:emlsig150}
\end{figure*}
\clearpage
\begin{figure*}
\begin{center}
	\includegraphics[width=1.0\textwidth]{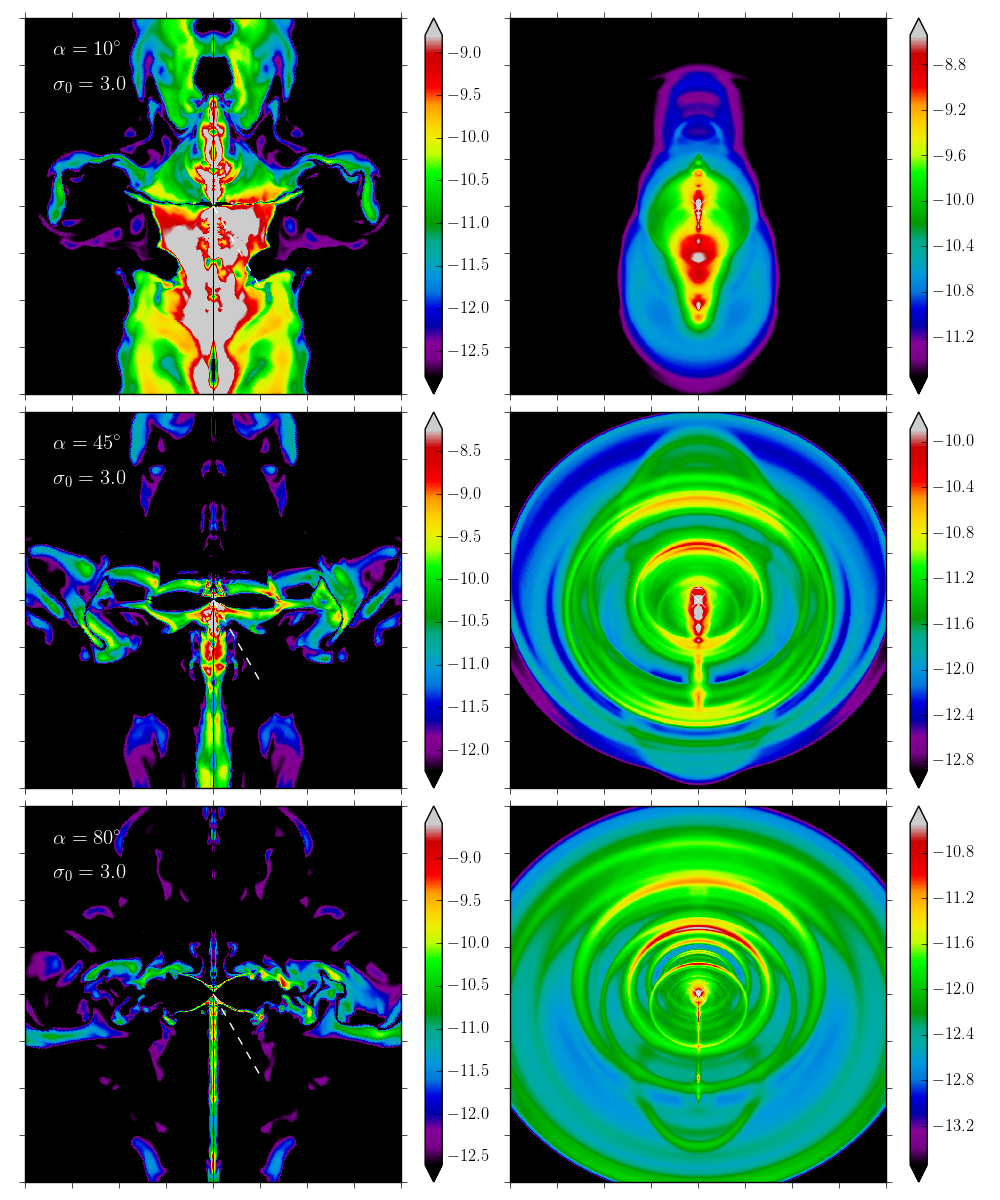}
 \end{center}
     \caption{As figure \ref{fig:emhsig}, but for a viewing angle $\theta_{view} = 150^{\circ}$.}
    \label{fig:emhsig150}
\end{figure*}
\clearpage
\end{document}